# Attoclock reveals geometry for laser-induced tunneling

Adrian N. Pfeiffer[1*], Claudio Cirelli[1], Mathias Smolarski[1], Darko Dimitrovski[2$], Mahmoud Abu-samha[2], Lars Bojer Madsen[2], Ursula Keller[1]

[1] Physics Department, ETH Zurich, 8093 Zurich, Switzerland
[2] Lundbeck Foundation Theoretical Center for Quantum System Research,
Department of Physics and Astronomy, Aarhus University, 8000 Aarhus C, Denmark

[*] Email: apfeiff@phys.ethz.ch; Telephone: +41 44 6332172; Fax: +41 44 633 1059
[$] Email: darkod@phys.au.dk; Telephone: +45 89423668; Fax: +45 8612 0740

**Summary**

Tunneling plays a central role in the interaction of matter with intense laser pulses, and also in time-resolved measurements on the attosecond timescale. A strong laser field influences the binding potential of an electron in an atom so strongly, that a potential barrier is created which enables the electron to be liberated through tunneling. An important aspect of the tunneling is the geometry of the tunneling current flow. Here we provide experimental access to the tunneling geometry and provide a full understanding of the laser induced tunnel process in space and time. We perform laser tunnel ionization experiments using the attoclock technique, and present a correct tunneling geometry for helium and argon. In addition for argon the potential barrier is significantly modified by all the electrons remaining in the ion, and furthermore from a quantum state whose energy is Stark-shifted by the external field. The resulting modified potential geometry influences the dynamics of the liberated electron, changes important physical parameters and affects the interpretation of attosecond measurements. These effects will become even more pronounced for molecules and surfaces, which are more polarizable.

One of the most striking manifestations of the rules of quantum mechanics is the possibility for a particle to move from one side of a potential barrier to the other regardless of the energy height of that barrier. This includes the classically forbidden case, referred to as tunneling, where the potential energy of the barrier is higher than the kinetic energy of the particle. In the research area of strong laser field interactions and attosecond science[1], tunneling of an electron through the barrier formed by the electric field of the laser and the atomic potential [Fig. 1a] is typically assumed to be the initial key process that triggers subsequent dynamics[1-3]. In linearly polarized laser fields electron tunneling is expected to eventually lead to above threshold ionization, enhanced double ionization and coherent emission up to the X-ray regime with high-order harmonic generation (HHG)[4-7]. Therefore a detailed understanding of the tunneling step is of paramount importance for attosecond science including generation of attosecond pulses[8,9] and attosecond measurement techniques[10-12].

To fully understand the electron tunneling process we need to know the tunneling geometry, the transition probability or rate, the time it takes to tunnel through the laser-induced barrier, and the momentum distribution of the electron at the exit of the tunnel. In this work, the tunneling geometry is the spatial geometry of the tunneling current flow for field-induced tunneling in atomic/molecular systems. New attosecond streaking techniques[10,11] have been extremely useful to characterize this tunneling process. For example, laser-induced tunneling from atoms revealed a step-like tunneling rate, confirming the picture of bursts of electrons being emitted every half cycle[13]. The technique was based on attosecond streaking[10] where a short attosecond pulse is used to remove an electron from a bound state and a time-dependent laser field is used as a time reference. The moment the electron exits from the bound state it is accelerated by this laser field, which is referred to as the streaking field. The final energy and momentum of the liberated electron, which are used to infer time-dependence in the ionization dynamics, depends on the exact exit point in space and time, typically given by the initial tunneling step. The question of tunneling time in laser-induced tunneling was experimentally addressed using the attoclock technique[11,14], and a zero tunneling time with an upper bound of approximately 10 as was reported.

Here we take these investigations a crucial step further and use the attoclock technique to obtain the first experimental information about the electron tunneling geometry and exit point. We confirm vanishing tunneling delay time, show the importance of the inclusion of Stark shifts and report on multielectron effects clearly identified by comparing results in argon and helium atoms. Our combined theory and experiment allows us for the first time to single out the geometry of the inherently 1D tunneling problem, through an asymptotic separation of the full 3D problem [Fig. 1]. The emerging physical picture is referred to as TIPIS - **T**unnel **I**onization in **P**arabolic coordinates with **I**nduced dipole and **S**tark shift. Our findings have implications for laser tunnel ionization in all atoms and in particular in larger molecular systems with correspondingly larger dipoles and polarizabilities.

**Attoclock and experiment**
The attoclock technique was described in detail elsewhere[11,14]. In brief, the attoclock is also an attosecond streaking technique as described above but here the time reference is given by a close-to-circularly polarized laser field. The rotating electric field vector gives the time reference similar to the hands of a clock, and the time is measured by counting fractions of cycles with the exact angular position of the rotating electric field. In this way it is possible to obtain attosecond time resolution by employing a femtosecond pulse. The attoclock was used to set an upper limit to the tunneling delay time during the tunnel ionization process in helium[14]. In this case a very short few femtosecond pulse was used to both ionize the helium and to provide the time reference. The pulse duration was sufficiently short such that the ionization event is limited to within one optical cycle around the peak of the pulse.

Because of the close-to-circular polarization, no re-scattering of the liberated electron with the parent ion occurs and the ionization event is very well isolated. After the tunneling event, the liberated electron is considered to propagate classically in the combined laser field and the potential of the parent ion, so that the instant of ionization can be mapped to the angle of the final momentum of the electron in the polarization plane, measured with cold target recoil ion momentum spectroscopy[15] [Fig. 2]. Depending on data acquisition time, an accuracy of a few attoseconds can be achieved because the measurement is based on a peak search for the exact angle at which the highest electron count rate is observed.

In this work, the attoclock technique is used to measure the *offset angle* $\theta$ [Fig. 3] which is directly related to the complex parent ion interaction assuming no measurable tunneling delay time. This offset angle is a extremely sensitive parameter for the detailed tunneling geometry information. The laser field alone in the so-called strong field approximation (SFA)[16-18] generates a maximum of the final electron momentum distribution along the minor axis of the polarization ellipse describing the close-to-circular polarization of the laser field (see online material). The Coulomb correction to the SFA has been a highly discussed topic[19-22]. The attoclock cycle, the time zero (i.e. $\theta = 0$), and the exact time evolution of the streaking laser field are fully characterized independently (online material). We minimize systematic errors in the angular streaking using both clockwise and anti-clockwise polarized pulses [Fig. 3], which we apply to the attoclock for the first time in this study. The ionization event then remains the same but the clockwise and anti-clockwise laser fields will streak the electron at the exit of the tunnel by equal amounts but in opposite directions, $\theta$ being the average of the two resulting angles.

In the experiments, we vary the peak intensity and therefore the Keldysh parameter[16] from 0.5 to 1.1 and to 1.4 for helium and argon, respectively. The laser focus and atom density were adjusted such that the number of ionization events per shot is much smaller than one. Figures 4 and 5 display the angular shift $\theta$ due to the interaction with the ionic potential during the angular streaking, as explained in Fig. 3. No significant intensity dependence of $\theta$ is observed for helium over the investigated intensity range [Fig. 4], while argon exhibits a monotonic downwards trend of $\theta$ with increasing intensity [Fig. 5].

**Theory**
Time-dependent Schrödinger equation (TDSE) simulations[23] are in agreement with our experiments [Fig. 5b] but could only be performed over a limited intensity range without serious numerical problems. These problems are caused by the close-to-circular polarization of the field requiring a full 3D-calculation, and the very high intensity requiring a high number of angular momenta and magnetic quantum numbers to account for the very many photon exchanges. Therefore we use simpler models to capture the essential physics.

The conventional models in strong-field physics are the strong-field approximation (SFA)[16-18] and the tunneling model[24] [Fig. 1]. The former does not account for the interaction of the outgoing electron with the remaining ion and predicts[23] vanishing $\theta$ in disagreement with experiment. The standard semi-classical tunneling model, consists of an initial tunneling step and subsequent classical propagation of the electron trajectory starting at the outer turning point of the tunnel [Fig. 1]. The classical trajectory is determined by the electric field of the remaining laser pulse, but significant corrections are induced by the ionic potential and the Coulomb correction from the ionic potential has been discussed intensively[19,20]. The impact of the ionic potential is more important in the vicinity of the parent ion, but the close-to-circular polarization steers the electron away and hence any deviation from the laser-field driven kinematics can be attributed directly to the ion potential.

The angle of the electron momentum is especially sensitive to the ion-electron interaction: The attraction to the ion at the beginning of the electron trajectory is reflected in an angular offset θ compared to the laser-only trajectory which remains unchanged as the laser field further evolves [Fig. 2b and Fig. 3]. Without the additional streaking forces from the parent ion (i.e. Coulomb force, polarizability etc.) the angular offset θ would be zero[14,23], and the final momentum would then be given by the integral of the electric field from the instant of ionization $t_0$, i.e., the final momentum would follow the vector potential $\mathbf{A}(t_0)$ [Fig. 3][25]. Even without considering the effect of the ionic potential, any tunneling delay time will manifest itself as an angular offset, because instead of $\mathbf{A}(t_0)$, the final momentum would then be given as $\mathbf{A}(t_0 - t_{tun})$, where $t_{tun}$ would be the tunneling delay time. The offset angle, monitored in this experiment, is extremely sensitive to the tunneling delay time. One degree in θ corresponds to $t_{tun}$ ~ 5-10 attoseconds. The excellent agreement of our theory with our measurement confirms the previous results[14] of zero tunneling time within the experimental accuracy of 10 as.

Although the electron feels the full 3D potential [Fig1a], only the potential in the direction of the field is usually considered[26]. We will refer to this model as "field-direction model". Using that model, one can predict that over-the-barrier ionization (OBI) occurs for argon within the intensity range investigated in this work. The comparison between theory and experiment (Figs. 4 and 5b) shows that calculating the dynamics based on tunnel-exit points obtained from the field-direction model fails to reproduce the experimental trend in θ. The reason is that the problem at hand is non-separable in these coordinates and the tunneling process, which is per definition 1D, is therefore not correctly described.

A separation of the Schrödinger equation for the electron in the external field is only possible for the pure Coulomb problem, i.e. for hydrogen-like systems, where the separation is accomplished in parabolic coordinates ξ = r+z and η = r-z [see Fig 1b]. In that case the resulting one-dimensionality of the separated problem enables an analytical treatment of tunneling[27]. In the present case we consider multielectron systems, and show that a separation of the problem is still possible in position space at the ξ and η values of relevance for tunneling. In the inner region [Figs 1b and 1c], we do not know the potential, but our procedure to determine the exit points is not sensitive to this lack of knowledge (online material), as the regions of space where the exit points are, is separable in parabolic coordinates [Fig1c]. Our procedure reveals the universal tunneling geometry for atoms: in this respect the pure hydrogen atom and other atomic systems do not differ (online material).

The starting point in the derivation of the TIPIS model is the time-independent Schrödinger equation with a static field **F** on. Since the electric field of the laser pulse is slowly varying with respect to the electronic dynamics, a quasi-static approximation is justified. The outer electron that is going to be ejected in the tunneling process is slower than the inner electrons, allowing an adiabatic decoupling of the motion with a Born-Oppenheimer-like ansatz[28]. The resulting one-electron equation for the outer electron[29] reads (atomic units are used throughout the paper),

$$-I_p(F)\Psi = \left(-\frac{1}{2}\Delta + V_{ef}(\mathbf{F},\mathbf{r})\right)\Psi, \quad V_{ef}(\mathbf{F},\mathbf{r}) = -\frac{1}{r} - \frac{\alpha^I \mathbf{F}\cdot\mathbf{r}}{r^3}, \quad (1)$$

where $I_p(F) = I_p(0) + \frac{1}{2}(\alpha^N - \alpha^I)F^2$ is the Stark-shifted ionization potential[30-32], $V_{ef}(\mathbf{F},\mathbf{r})$ is the effective potential for atoms (valid at large and intermediate distances)[29], $\alpha^N$ is the static polarizability of the atom and $\alpha^I$ of its ion. The influence of the inner electrons on the outer electron dynamics is included through the polarizability of the ion[33], which is the main multi-electron effect that we derive consistenly in the model[29], and which is beyond a standard single-active electron model.

Equation (1), containing the multi-electron effect through the induced dipole, does not separate in parabolic coordinates. However separation in parabolic coordinates is possible in the limit, $\xi/\eta \ll 1$, which is satisfied under the present experimental conditions (online material). The induced dipole part of the potential only contributes to the equation for the η-coordinate, which becomes

$$\frac{d^2 f(\eta)}{d\eta^2} + 2\left(-\frac{I_p(F)}{4} - V(\eta,F)\right)f(\eta) = 0 \quad (2)$$

where the effective tunneling potential reads (Fig. 1c)

$$V(\eta,F) = -\frac{(1-\sqrt{2I_p(F)}/2)}{2\eta} - \frac{1}{8}\eta F + \frac{m^2-1}{8\eta^2} + \frac{\alpha^I F}{\eta^2}. \quad (3)$$

The exit points are determined by equating the potential and the energy term in equation (2) with $m = 0$. The tunnel exit points obtained from the separated problem (3) are generally larger than those obtained in field-direction model [Fig. 5a], and the inclusion of the Stark shifts (larger binding energy) and induced dipole of the ion (larger barrier) pushes the exit points even further away from the origin [Fig. 5a]. Since we separate the tunneling problem in 1D, the tunneling geometry is a line along the parabolic η coordinate but it defines a whole region in the 3D space whereas in the field direction model, the tunneling geometry is a line in a 3D space.

The offset angle θ, obtained using semi-classical models, is very sensitive to a change in the tunnel exit points [Fig. 5a], especially when intensities approach the OBI regime. The induced dipole and Stark shifts lead to an increase of the OBI [Fig. 5]. The experimental parameters are such that in the parabolic-separated problem, no over-the-barrier ionization occurs [Fig. 5a]. Because of the very small polarizabilities of He and He$^+$, the Stark shifts and the induced dipole barrier modification are negligible, and it is the separation in parabolic coordinates that leads to a plateau in the θ dependence on intensity, rather than the weak monotonic increase produced by the field-direction semi-classical model [Fig. 4]. For the case of argon, with larger polarizabilities, the inclusion of Stark shifts, and even more, the multi-electron effect through the increase of the barrier due to the ionic induced dipole, become decisive for the decrease of θ with increasing intensity. In contrast, the semi-classical model using tunnel exit points obtained in the field-direction model exhibits a non-monotonic dependence on laser intensity, having a pronounced local maximum close to the OBI intensity [Fig. 5b].

In addition to the strengths of the present model described above, we stress that the prediction for the trend in θ is insensitive to the exact form of the rate. The only requirement for obtaining the trend in θ is to have a rate that increases with increasing intensity (online material). This is advantageous since rates for systems different from the simplest atoms are difficult to obtain, and gives further confidence to the determination of the calculated exit points in time and space.

**Conclusions and perspectives**

The attoclock technique allowed us to gain new insights in laser-induced tunneling, one of the paradigms of modern strong-field and ultrafast science. With the offset angle θ in our attoclock experiments we are able to obtain direct access to the electron-ion interaction and the tunneling exit point. Specifically, this study confirmed a vanishingly small tunneling time and above all, revealed the natural geometry of laser induced tunneling in atoms, which allowed for an asymptotic separation in parabolic coordinates and reducing the non-separable 3D problem to a true 1D tunneling problem. Furthermore, our model showed very clearly by comparing the data for helium and argon that more contributions in the electron parent-ion interactions are important such as the Stark shifts of the bound-state energy levels and the multi-electron effect describing the action of the induced dipole of the core electrons on the liberated electron. These two effects give rise to a modification of the potential barrier in argon and a significant change of the tunnel exit point.

The extent to which the multielectron effects influence the tunneling dynamics is system-dependent as shown here by the difference in helium and argon. Argon is much easier to polarize than helium and is therefore affected more strongly. These new insights into the physics of laser tunnel ionization are universally present in all atomic and molecular systems and are rationalized in our TIPIS model. The model can be used to describe attosecond ionization dynamics for more complex targets in a tunneling regime or with large wavelengths lasers, situations in which TDSE methods cannot presently provide an answer.

Our experimental and theoretical investigations will have a high impact on attosecond sience and strong-field physics. They can be summarized as follows:

First, the intensity range where the electron still emerges below the tunnel barrier is significantly extended towards higher intensities. This is important for novel attosecond measurement concepts because the liberated electrons in the laser-induced tunneling regime exhibit more precisely defined properties. This enables time-dependent structure information when these electrons are then recollided and diffracted at the original target[34]. In addition theoretical tools suited for below the barrier ionization can now be used over a much larger intensity range.

Second, the multi-electron effects identified here will greatly affect further studies on larger molecules and on surfaces. Much less is known but recent evidence has emerged suggesting that the initial tunneling step needs to be revisited: multiple orbital tunneling[35-37], and field-induced shifts of the energy levels in polar molecules[32]. Larger molecules are much more polarizable then the noble gas atoms studied here, and the effects reported here will be visible especially in experiments employing circularly or near circularly polarized laser pulses that isolate the ionization event, as done in Holmegaard *et al.*[32]

Third, attosecond measurements typically rely on streaking techniques which are highly sensitive to the parent ion interaction[38,39]. If this interaction is not understood correctly, wrong conclusions could be drawn on possible time delays. To date attosecond streaking has been applied to atomic[10,40,41], molecular[42] or solid target[43]. The near infrared femtosecond pulse will have the potential to polarize the core and consequently lead to additional force terms like those identified in the present work.

**Supplementary Information** is linked to the online version of the paper.

**Acknowledgements**
This work was supported by the NCCR Quantum Photonics (NCCR QP) and NCCR Molecular Ultrafast Science and Technology (NCCR MUST), research instruments of the Swiss National Science Foundation (SNSF), and by ETH Research Grant ETH-03 09-2 and an SNSF equipment grant and by the Danish Council for Independent Research, Natural Sciences.

**Author Contributions**
A. N. Pfeiffer, C. Cirelli, M. Smolarski and U. Keller performed experiments and simulations. D. Dimitrovski, M. Abu-samha and L.B. Madsen developed the theory. All authors participated in the writing of the paper.

-

**Figure Legends**

**Fig. 1: Strong field tunneling**

Laser-induced adiabatic tunnelling picture (PPT[24] and Keldysh[16]) where the laser frequency is much smaller than the oscillation frequency of the bound-state electron. Our helium and argon experiments in the near infrared fulfil this approximation (see online material). This means that the combined 3D potential field of the ion and laser shown in (a) for y = 0 changes slowly. The only escape is then through tunnelling and a quasi-static potential barrier can be used. The field F is assumed to be pointing in the positive z direction. A cut of the potential along the instantaneous field direction, along z and for x=0 gives the field-direction model as given by the inset in (a).

(b) Illustration of parabolic coordinates which separate the Schrödinger equation for the electron in the combined Coulomb and laser potentials. The figure shows a cut for y = 0 and the red dashed curves correspond to contours with constant $\xi$, and the blue curves correspond to contours with constant $\eta$, the values of $\xi$ and $\eta$ are given next to the contours. The transparent magenta circle illustrates the approximate region in position space where the exact atomic potential is non-separable in parabolic coordinates and where the far-field expression for the dipole-terms (equation (3)) cannot be used. In (b) we see that when we vary $\eta$, the coordinate in which tunnelling occurs (see (c)), many values of x and z are probed and this effect obtained in the parabolic coordinates that separate the Schrödinger equation is not captured by the field-direction model.

(c) and (d) show the effective potential plus the binding energy for argon at the experimental intensity of $5\times10^{14}$ W/cm$^2$ along $\eta$ and along $\xi$, respectively.

The potential V(η,F) has a barrier through which the electron may tunnel, while the potential V(ξ,F) does not form a barrier and there is no possibility for the electron to escape by tunneling. The approximate expectation value of ξ for the wave function bound in V(ξ,F) is 1. With this expectation value of ξ, the full lines in (c) starting at η = 10 a.u. corresponds to ξ / η << 1. In this region separation in parabolic coordinates is possible and this is where the tunnel exit point is located. The dashed curves give the potential in the inner, nonseparable region where we assume that the potential forms a barrier. The red dashed line corresponds to the potential (3), and the black dashed line corresponds to the same potential with the induced dipole term multiplied with exp(-3 / η) to account for the fact that pure dipole-like potential can be used only in the far field.

**Fig. 2: Attoclock principle**

**(a) Attoclock[11,14] setup with a COLTRIMS (COLd Target Recoil Ion Momentum Spectroscopy)[15] apparatus:** An intense infrared (IR) laser pulse (2 cycles, 7 fs pulse duration, center wavelength 740 nm, 1 kHz pulse repetition rate) from a Ti:sapphire laser system with a two-stage filament compressor[44] is propagated through a quarter wave plate to produce a close-to-circularly (ellipticity 0.78) polarized laser field. The rotating electric field vector gives the time reference, similar to the hands of a clock. This pulse is then focused into a supersonic gas target to tunnel ionize helium or argon. The ions and electrons are guided onto time and position sensitive detectors, to determine, in coincidence, the momenta of the atomic fragments after the laser pulse. Much less than one ionization event occurs per pulse.

**(b) Angular streaking with both clockwise and anti-clockwise close-to-circular polarization:** The green arrow illustrates the laser propagation direction and the blue arrow represents the rotating electric field vector (i.e. the hand of the attoclock) at peak intensity. The time evolution of the peak of the electric field vector is shown in orange for both clockwise and anti-clockwise polarization. The tunneling delay time[14] is defined by the angular difference between the maximum of the electric laser field (i.e. blue arrow), which induces the highest tunneling ionization rate, and the direction of the laser field when the electron exits the tunnel. At the exit of the tunnel, the electron is assumed to be in the continuum and to experience the acceleration of the strong laser field present at that moment. Depicted are electron trajectories[25] in red and black, with and without the interaction with the parent ion respectively, which then determine the momenta at the target after the pulse [Fig. 2a].

**Fig. 3: Systematic error reduction in angular streaking**

Applying both clockwise and anti-clockwise polarization under the same experimental conditions minimizes systematic errors with the attoclock. The dynamics of the tunnel ionization event is preserved, the clockwise and anti-clockwise field will streak the electron at the exit of the tunnel by equal amounts but in opposite directions.

(a) The angle $\alpha$ between the fast axis of a quarter wave plate and the polarization plane of the incoming laser pulse determines the ellipticity $\varepsilon$ and the angle of the polarization ellipse $\beta$. The measurement (red) is in excellent agreement with the simulation (black)[25]. For two angles α separated by 90°, the polarization is identical, except for the turning direction of the electric field vector.

(b) The ion momentum distribution peak (dashed white line) rotated by 90° - compared to the major polarization axis at angle β due to the propagation in the laser field alone[25]. An additional angular offset θ into the electric field turning direction is observed due to the interaction with the ion. The intensity dependence of θ is shown in the lower graph for anti-clockwise and clockwise turning fields. For all our results as presented in Fig. 4 and 5b, we considered both measurements from anti-clockwise and clockwise (with sign change in θ) streaking fields. Wavelength and pulse duration are given in the caption of Fig. 2.

**Fig. 4: Offset angle for helium**

Angular offset θ, obtained from ion momentum distributions, measured in Helium as a function of laser intensity, shown together with different model curves (see text for notations). Wavelength and pulse duration are given in the caption of Fig. 2. The label "Cartesian coord" refers to the field-direction model. The label "Parabolic coord" refers to the parabolic separated model without inclusion of the Stark shifts and multielectron effects of equation (2), "Parabolic+Stark shift" refers to the problem of eqtaion (2) with inclusion of Stark shifts, but without inclusion of multielectron efects. The label "TIPIS" refers to our full TIPIS model.

**Fig. 5: Offset angle for Argon**

(a) Exit of the tunnel as a function of electric field strength and laser intensity for different models. If the field-direction model is used, OBI is reached at about $4 \times 10^{14}$ W/cm$^2$. Over this intensity, the electron is placed at the combined potential saddle point with an initial nonzero velocity (displayed as dashed line) longitudinal to the electric field at the instant of ionization (online material). When parabolic coordinates are used the dynamics is tunneling below the barrier and trajectories are launched from the tunnel exit with zero initial velocity. The figure shows that the separation in parabolic coordinates, the Stark shift of the initial state, and the induced dipole contribution of the bound electrons all significantly influence the starting point of the classical trajectory.

(b) Experimental data of θ, obtained from electron momentum distributions in argon, as a function of laser intensity, together with the curves predicted by the TDSE and different models. For all the models, the exit of the tunnel and the eventual initial velocity is determined as in (a). The measured trend is only reproduced when the multi-electron effect due to the polarized ion is included. Wavelength and pulse duration are given in the caption of Fig. 2. The labelling of the different curves is explained in the caption of Fig. 4.

# Figures

## Fig. 1

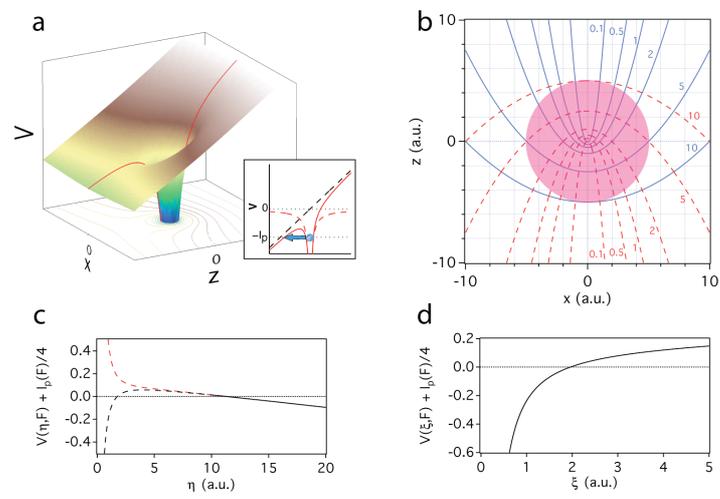

## Fig. 2

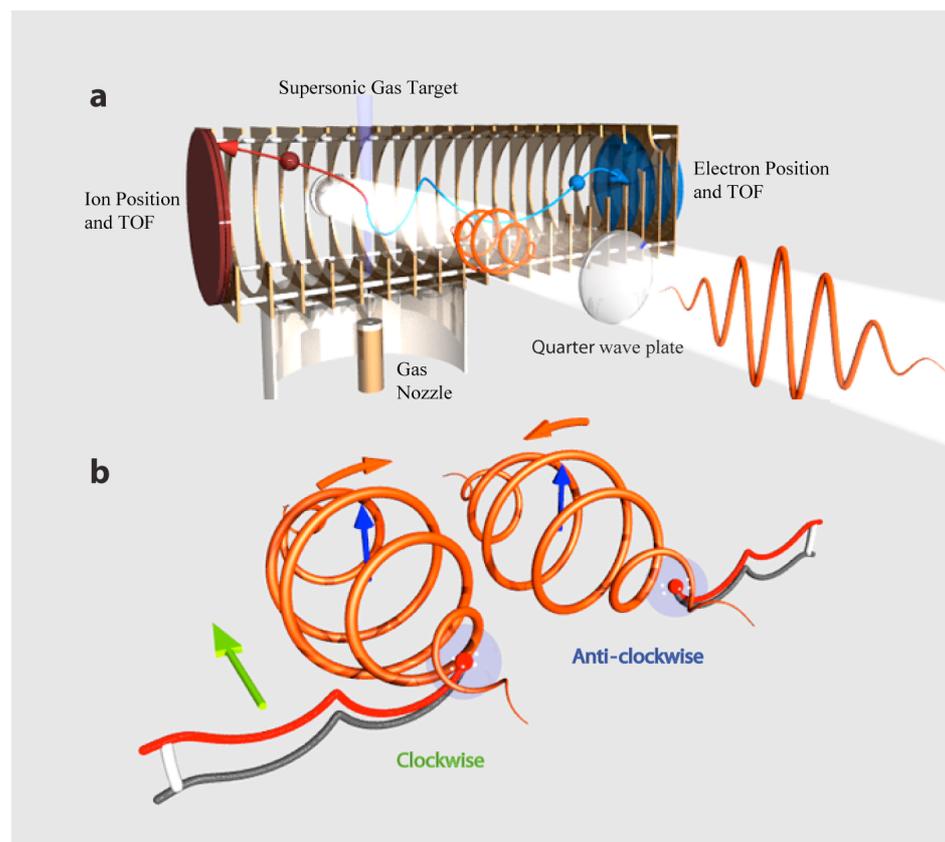

**Fig. 3**

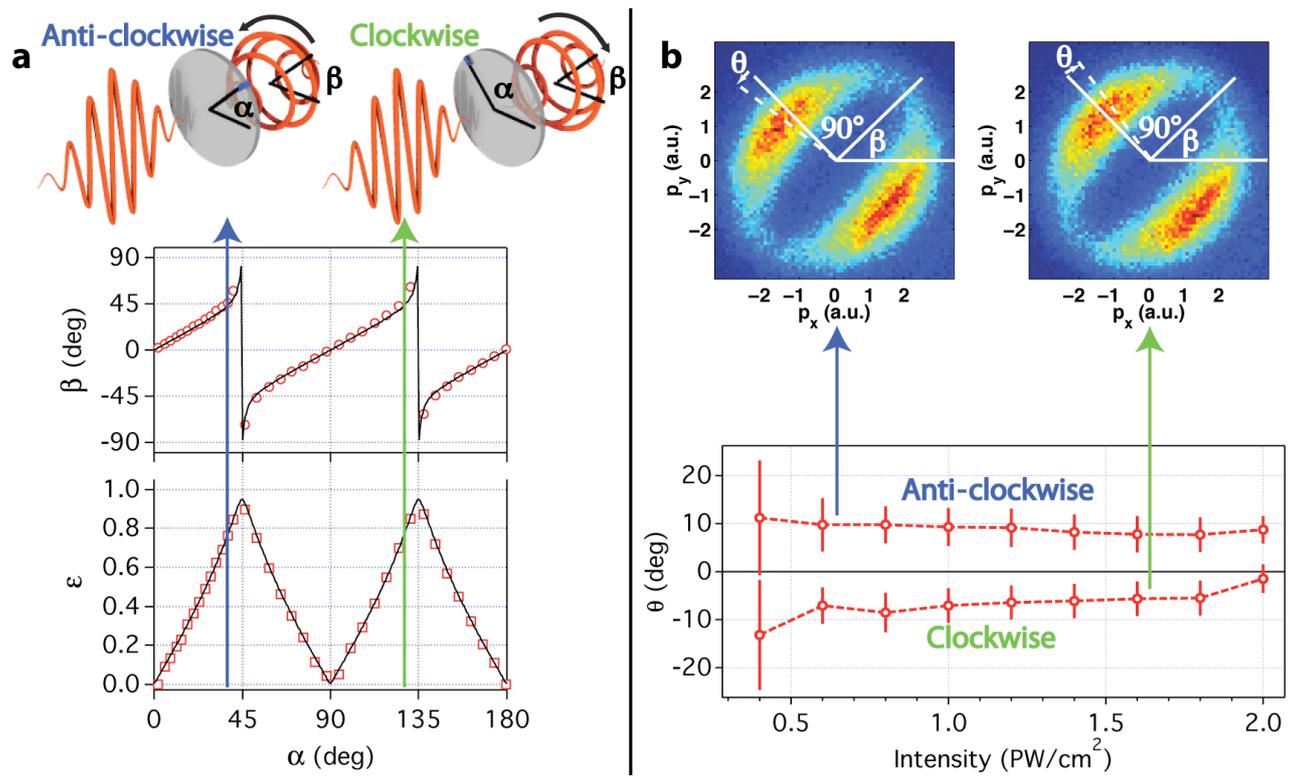

**Fig. 4**

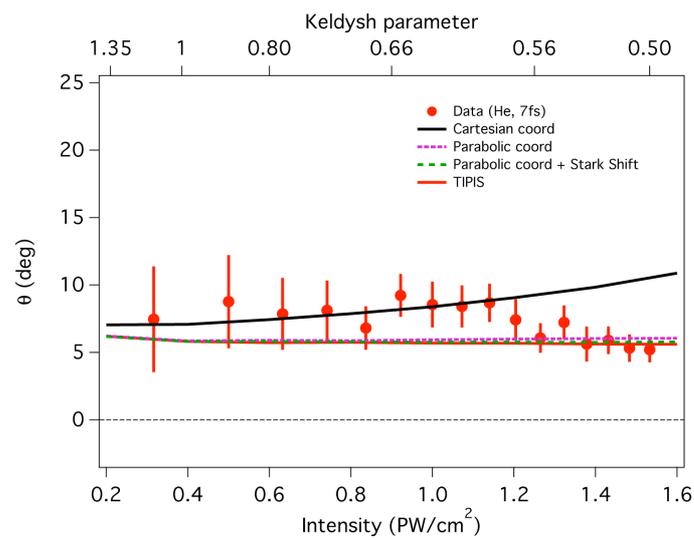

**Fig. 5**

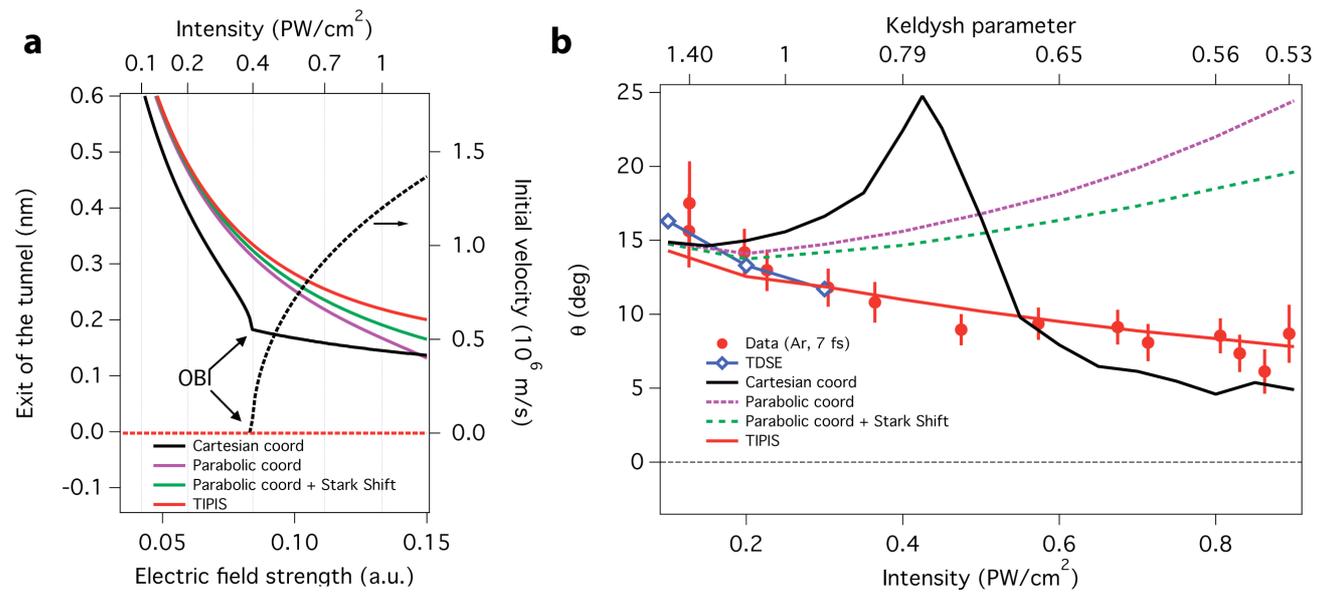